# Highly transparent contacts to the 1D hole gas

# in ultra-scaled Ge/Si core/shell nanowires

M. Sistani[1*], J. Delaforce[2*], R. Kramer[2], N. Roch[2],

M.A. Luong[3], M.I. den Hertog[2], E. Robin[3], J. Smoliner[1], J. Yao[4], C.M. Lieber[5,6], C. Naud[2],

A. Lugstein[1], O. Buisson[2#]

[1] Institute of Solid State Electronics, TU Wien, Gußhausstraße 25-25a, 1040 Vienna, Austria

[2] Université Grenoble Alpes, CNRS, Institut NEEL UPR2940, Grenoble, France

[3] Université Grenoble Alpes, CEA, IRIG-DEPHY, F-38054 Grenoble, France

[4] Department of Electrical and Computer Engineering, Institute for Applied Life Sciences, University of Massachusetts, Amherst, Massachusetts, 01003, USA.

[5] Department of Chemistry and Chemical Biology, Harvard University, Cambridge, Massachusetts, 02138, USA

[6] School of Engineering and Applied Science, Harvard University, Cambridge, Massachusetts, 02138, USA

KEYWORDS: nanowire heterostructure, germanium, 1D hole gas, superconductor-semiconductor hybrids, hard superconducting gap, transparent contacts

*These authors contributed equally

ABSTRACT

Semiconductor-superconductor hybrid systems have outstanding potential for emerging high performance nanoelectronics and quantum devices. However, critical to their successful application is the fabrication of high quality and reproducible semiconductor-superconductor interfaces. Here, we realize and measure axial Al-Ge-Al nanowire heterostructures with atomically precise interfaces, enwrapped by an ultra-thin epitaxial Si layer further denoted as Al-Ge/Si-Al nanowire heterostructures. The heterostructures were synthesized by a thermally induced exchange reaction of single-crystalline Ge/Si core/shell nanowires and lithographically defined Al contact pads. Applying this heterostructure formation scheme enables self-aligned quasi one-dimensional crystalline Al leads contacting ultra-scaled Ge/Si segments with contact transparencies greater than 98%. Integration into back-gated field-effect devices and continuous scaling beyond lithographic limitations allows us to exploit the full potential of the highly transparent contacts to the 1D hole gas at the Ge-Si interface. This leads to the observation of ballistic transport as well as quantum confinement effects up to temperatures of 150 K. Low temperature measurements reveal proximity-induced superconductivity in the Ge/Si core/shell nanowires. The realization of a Josephson field-effect transistor allows us to study the subharmonic energy-gap structure caused by multiple Andreev reflections. Most importantly, the absence of a quantum dot regime indicates a hard superconducting gap originating from the highly transparent contacts to the 1D hole-gas, which is potentially interesting for the study of Majorana zero modes. Moreover, underlining the importance of the proposed thermally induced Al-Ge/Si-Al heterostructure formation technique, our system could pave the way for novel key components of quantum computing such as gatemon or transmon qubits.

The capability to customize the morphology and size of low-dimensional nanostructures, such as nanowires (NWs), and thus tune the associated electronic and optical properties, has triggered substantial research interest.[2,3,4] Especially band-structure engineering by controlled epitaxial growth of core/shell NWs provoked the investigation of 1D hole-gas systems,[4] attractive for fundamental studies of low-dimensional transport as well as future high-performance nanoelectronic or quantum devices.[5,6,7,8,9] Further, heterostructures of dissimilar materials with unique structure-property relationships and interactions originating from the contributions of individual low-dimensional components may enable novel electronic or photonic devices that are outclassing or even unattainable for planar geometries.[10,11] However, fabricating interconnects is a crucial step towards the integration of such future ultra-scaled devices and requires sophisticated nanostructure formation techniques and precise lithography. To overcome these limitations, material combinations with no intermetallic phase formation, such as the Al-Ge system, enabling true metal-semiconductor heterostructures with abrupt metal-semiconductor interfaces received a considerable amount of attention.[12,13,14,15,16,17] Within this paper, we apply quasi 1D superconducting Al leads to ultra-scaled Ge/Si core-shell channels forming highly transparent contacts to the 1D hole gas. Such a monolithic superconductor-semiconductor heterostructure enables an exchange of Cooper-pairs between two highly transparent superconducting contacts through the hole gas. This leads to supercurrent induced by the superconducting proximity effect[18] which is an important prerequisite for a Josephson field-effect transistor (JoFET). That could be integrated into a gate-tunable superconducting qubit, often referred to as a gatemon.[19] Further, this architecture due to the strong spin-orbit coupling of holes in Ge could be an attractive candidate for the study of Majorana zero modes and development of topological superconducting qubits.[16,20]

Our approach for the synthesis of a monolithic Al-Ge/Si-Al NW heterostructure featuring highly transparent contacts to a 1D hole gas is based on vapor-liquid-solid[21] grown core/shell NWs with a Ge NW core diameter of 30 nm and a Si-shell thickness of about 3 nm. For such NWs contacted by Al pads we demonstrate the realization of tunable Ge segment lengths by the selective substitution of the Ge core by crystalline Al (c-Al) utilizing a thermally induced exchange reaction while maintaining the ultra-thin semiconducting shell wrapped around (figure 1a).[13] The insets of figure 1a show energy dispersive X-ray spectroscopy (EDX) chemical maps at the respective positions and proves an intact Si shell around the Ge NW segment and the Al NW. The monolithic heterostructure formation enables self-aligned quasi-1D c-Al leads contacting the Ge/Si segment. A SEM image of the actual heterostructure arrangement is shown in figure 1b. For an analysis of the Al-Ge interface, high-resolution high angle annular dark field (HAADF) scanning transmission electron microscopy (STEM) oriented along the $[11\bar{2}]$ direction of observation of the Ge crystal was performed on a probe corrected FEI Titan Themis working at 200 kV (obtained at the cyan dashed square shown in figure 1b). All investigated Al-Ge interfaces showed a very sharp metal-semiconductor interface, and we observed an epitaxial relation between the exchanged c-Al and Ge part along the Ge(111) growth plane, without the visibility of any crystal defects. An intensity profile obtained in the HAADF STEM image (figure 1c), shows that going from the Ge segment to the converted c-Al region the intensity changes over about 1 nm. However, if we measure the local lattice spacing, as shown in the inset, we find the last plane of higher HAADF STEM intensity has the lattice spacing close to the value of a Ge (111) plane (about 0.32 nm). The final Ge (111) plane is directly followed by a lower intensity plane with the lattice spacing close to the value expected for an Al (111) plane (0.24 nm), indicating a perfectly abrupt interface. Using this Al metallization scheme and consecutive annealing steps, the Ge/Si channel length can be tuned by the Al-Ge exchange procedure beyond lithographic limitations down to 10 nm (see supporting figure S1).

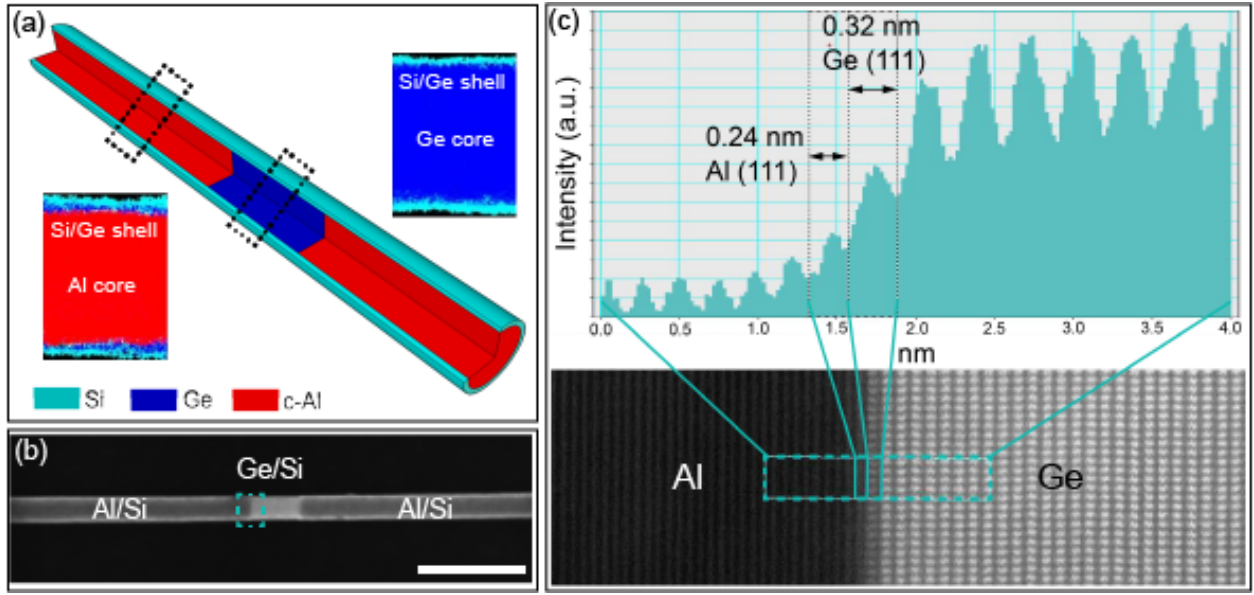

**Figure 1.** *(a) Schematic illustration of an axial Al-Ge-Al NW heterostructure with an ultra-thin semiconducting shell wrapped around. The insets show EDX chemical maps at the respective positions along the heterostructure indicating an intact semiconducting shell around the entire Al-Ge-Al heterostructure. (b) SEM image of the actual heterostructure arrangement. Scale bar is 200 nm. (c) High-resolution HAADF STEM obtained at the Al-Ge interface and corresponding intensity profile obtained at the cyan dashed square shown in (b).*

To investigate the electrical transport properties, the intrinsic Ge/Si core/shell NWs were integrated in back-gated FETs. A schematic illustration of the device architecture and the respective biasing is displayed in the upper inset of figure 2. The transport measurements were carried out using a two probe configuration; hereafter we have subtracted the wiring resistance. The main plot of figure 2 shows the comparison of room temperature I/V characteristics at gate-voltage ($V_G = 0$ V) of an intrinsic Ge NW (blue) and a Ge/Si core/shell NW (black) both with a physical transistor channel length of 2 μm. For the core/shell NW a clean confinement potential in the Ge core is expected[8,22] in view of the epitaxial growth of the Si-shell resulting in a high mobility 1D hole gas[4,23] (see lower inset of figure 2). The origin of the hole gas is associated with the abrupt discontinuity of the band-structure at the Ge-Si interface showing a band-offset of approximately 500 meV.[4] This causes holes to flow from Si to Ge to maintain a constant chemical potential throughout the arrangement. Consequently, the band edges are bent at the interface and holes are confined in the Ge close to the Ge-Si interface forming a hole gas.[22] Thus, in comparison with the intrinsic Ge NW the resistance of the Ge/Si core/shell NW is more than two orders of magnitude lower. The first annealing cycle reduces the channel length of the Al-Ge/Si-Al NW heterostructures from 2 μm (black) to $L_{Ge/Si}$ = 470 nm (green). We observed a moderate decrease in resistance of about 60%, which we dedicate to a combined effect of the reduction of the channel length and a change of the contact architecture from the Al pad atop of the Ge core to a quasi 1D monolithic Al-Ge contact. The second annealing cycle reduces the channel length further to $L_{Ge/Si}$ = 40 nm (red) resulting in a roughly tenfold reduction in resistance. In comparison with a ballistic bare Ge NW based device of the same channel length, the resistance of the Ge/Si core/shell heterostructure is by a factor of 5 lower.[24] Therefore, even though multiple consecutive annealing cycles were performed, Ge/Si core/shell NWs still reveal a 1D hole characteristic proving the thermal stability of the Si shell and the selective exchange of the Ge core.

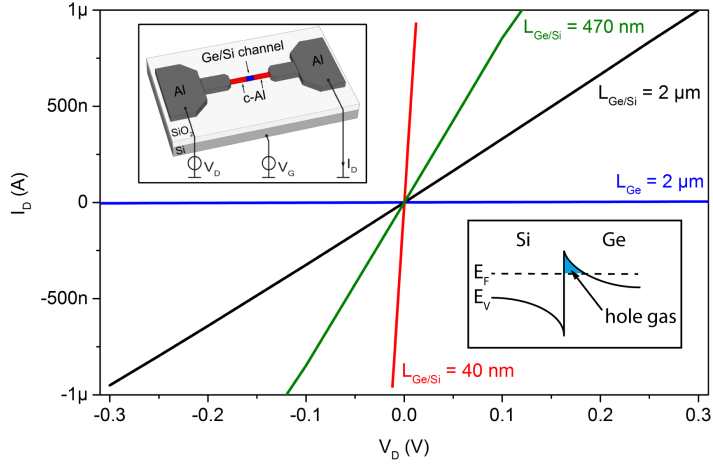

***Figure 2.*** *Comparison of the I/V characteristic of an intrinsic Ge NW (blue), a Ge/Si core/shell NW (black) and Al-Ge/Si-Al heterostructure device with varying channel lengths of $L_{Ge/Si} = 470$ nm (green) and $L_{Ge/Si} = 40$ nm (red) achieved by consecutive annealing steps. The upper inset shows a schematic illustration of the heterostructure NWs integrated in a back-gated FET configuration. A band diagram at the Ge/Si heterojunction in Ge/Si core/shell NWs is shown in the lower inset. $E_V$ and $E_F$ are the valence band edge and the Fermi energy, respectively.*

Figure 3a shows the conductance vs gate voltage (G-$V_G$) characteristic of the back-gated Ge/Si core/shell NW FET device with the channel length of $L_{Ge/Si} = 40$ nm measured in the temperature range from T = 5 K to 300 K. When the gate voltage is swept from positive to negative the conductance increases, thus the device behaves like a p-type accumulation FET device.[25] Cooling the sample does not change the overall conductance but at T = 150 K and even more pronounced for lower temperatures, distinctive plateau like features can be observed. As the channel length of this particular device is below the scattering mean free path in Ge/Si core/shell NWs of 70 nm,[26] we associate this quantization of conductance in steps of $G_0 = 2e^2/h$ with one-dimensional spin-degenerate sub-band-resolved quantum ballistic transport.[4,27,28] In the inset the resistance (R = 1/G) of the plateaus taken at the points marked by black arrows are plotted against the conducting channel number, associated with that plateau (n). Given that the Ge segment is in the ballistic limit and that we are in linear regime we can estimate the transparency of the interface by linearly fitting $R = (R_0 + R_I)/n$,[29] blue curve of inset, where $R_0 = h/2e^2$ is the quantized resistance and $R_I = R_0(1-T)/T$ is the interface resistance due to scattering. The slope of the linear fit gives a $R_0 + R_I$ value of $1.02R_0$ resulting in a transparency of approximately 98%. Furthermore, the G-$V_G$ measurement recorded at T = 5 K hints at the conductance anomaly at $0.7G_0$ (blue arrow), which is considered as an intrinsic low-temperature sub-$G_0$ feature of mesoscopic devices.[30]

To investigate the observed plateaus below the superconducting transition temperature of the Al contacts ($T_C = 1.19$ K)[31] measurements were carried out in a pumped $^3$He cryostat. Figure 3b shows traces of the bias voltage dependent differential conductance (G = $I_D$ / $V_D$) for fixed gate voltages ranging from $V_G = 30$ V to -30 V with a step size of 167 µV at T=450 mK. The conductance traces bunch into five thick lines of constant conductance, separated by regions of low trace density. Five conductance channels are clearly visible in figure 3b, with each bunching region occurring near an integer multiple of $G_0$. We attribute the dip in conductance around $V_D \approx \pm 4$ mV to be a consequence of the conductance evolving from integer plateaus, in the low bias regime, to 'half' plateaus, in the high bias voltage (nonlinear) regime.[8]

Considering the low temperature of the measurement (T < $T_C$), the conductance peak at zero-bias is a clear first indication for induced superconductivity in the Ge/Si core/shell segment. Further, an indication for quantization of these superconducting features is supplied by supporting figure S2, which shows an overlay of the differential conductance recorded above (T = 2 K) and below (T = 450 mK) $T_C$. The observation of quantized conductance further endorses our achievement of a near atomically precise interface between Al contacts and Ge/Si core/shell segments.

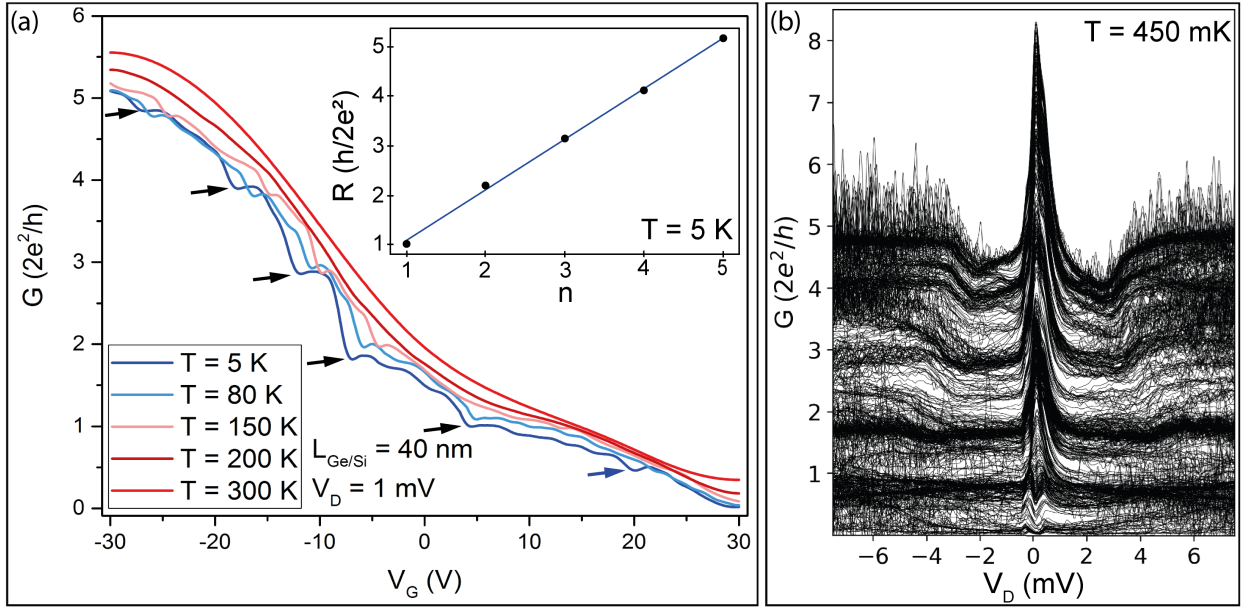

**Figure 3.** *(a) G-$V_G$ characteristics of the Al-Ge/Si-Al heterostructure device with a channel length of $L_{Ge/Si}$ = 40 nm measured at different temperatures between T = 5 K and 300 K. The conductance G was directly obtained from the measured current according to G = $I_D$ / $V_D$ and is plotted in units of $G_0$. The black arrows indicate the quantized conductance plateaus and the blue arrow indicates the 0.7$G_0$ plateau at 5 K. The inset shows the resistance (R) of the quantized conductance plateaus vs the conducting channel number (n). (b) G = $I_D$ / $V_D$ with series resistance of 370 $\Omega$ subtracted waterfall plot from $V_G$ = 30 V to -30 V in 167 $\mu$V steps measured at T = 450 mK.*

To investigate the high conductance regime for gate-voltages between -30 V and 0 V, we carried out current biasing measurements at T = 420 mK. The gate-voltage was swept from high to low and the bias from negative to positive. Figure 4a shows the differential resistance (d$V_D$ / d$I_D$) as a function of bias current and gate voltage between -30 V < $V_G$ < 0 V. The density plot clearly shows a region of near zero resistance centered around $I_D$ = 0 nA. This $V_G$ and $I_D$ dependent region suggests there is a regime where dissipationless supercurrent passes through the Ge/Si core/shell segment. To illustrate the $V_G$ dependence on the transport dynamics, the lower inset shows a resistance slice at $I_D = 0$ nA; a regime of high zero-bias resistance, of the order of several quantum of resistance, between $V_G = 0$ V and -3 V preludes a strongly decreasing resistance of the device as the gate voltage is decreased. Decreasing from $V_G = -3$ V, the zero-bias resistance of the heterostructure device drops abruptly and finally converges to a small finite resistance of 25 $\Omega$. We attribute this finite resistance to thermally activated phase slips. Such phase slips are of significance in Josephson junctions, when the Josephson energy ($E_J = \Phi_0 I_C / 2\pi \approx 5 \times 10^{-24}$ J) is of the order of the thermal energy ($k_B T \approx 6 \times 10^{-24}$ J).[32]

To further show the dependence of $I_C$ on the gate voltage induced electric field, the upper inset of figure 4a shows the I/V characteristic of the heterostructure device for particular gate-voltages, illustrating the ability to tune $I_C$. These observations clearly demonstrate that such a device, with a short channel, possess gate voltage mediated superconducting proximity effect.[32,33] As the Ge/Si core/shell heterostructure device shows characteristics of a p-type semiconductor, a more negative $V_G$ increases the number of conductance channels, resulting in higher $I_C$ and an increased conductance in the normal state. At approximately $V_G = -25$ V, the critical current saturates at a value of about $I_C = 15$ nA.

Figure 4b shows the respective plot in the gate-voltage range between $V_G = 0$ V and 30 V based on a measurement of differential conductance ($dI_D / dV_D$) as a function of bias voltage and gate voltage. The inset shows raw I/V curves for $V_G = 28$ V, 20 V, and 10 V, further illustrating the transport dynamics of the device. The density plot reveals that even for a high positive gate-voltage up to $V_G = 30$ V there are no Coulomb blockade effects, indicating the absence of a quantum dot regime. This is supported by the significant conductivity outside the superconducting gap of $V_G = 28$ V of approximately 0.3 $G_0$. This observation is in contrast to pure Si or Ge based devices featuring Schottky barriers[20] or Ge/Si core/shell based NW devices contacting the Si-shell,[35] forming a tunnel barrier at low temperatures. These findings further indicate that our Al-Ge/Si-Al heterostructure devices feature highly transparent contacts to the 1D hole gas, which is in good agreement with Xiang et al.[8] In the low conducting regime (25 V < $V_G$ < 30 V) we observe a superconducting gap with a minimum gap ratio of <$G_G$> / <$G_N$> = 0.03, where <$G_G$> is the average conductance inside the gap, across a $V_D$ range from -0.05 mV to 0.05 mV, and <$G_N$> the average conductance outside the gap, across a $V_D$ range from -0.7 mV to -0.6 mV.

Figure 4c shows a slice of the differential conductance with respect to $V_D$ in the low conducting regime ($V_G$ = 28 V, taken from figure 4b) and the high conducting regime ($V_G$ = -29.5 V, taken from supporting figure S3) at T = 420 mK. In both regimes, a family of peaks are observed symmetrically around $V_D$ = 0 V. In the high conducting regime, the peak at zero bias shows a differential conductance higher than 200 $G_0$ and corresponds to the "infinite conductivity" of the superconducting state of the Ge/Si core/shell channel. Further, the peaks at finite $V_D$ correspond to the subharmonic energy-gap structure caused by multiple Andreev reflections (MARs),[18] with peak positions given by $eV_n = 2\Delta/n$.[34] Such features arise from a progressive increase of the incident carrier energy as the carrier reflects between the two interfaces and thus mark Andreev channels present in superconductor – normal – superconductor (S-N-S) junctions for applying bias voltages below the superconducting gap. Taking the superconducting gap ($\Delta$) to be half of, $V_D$ = 0.37 mV, the position of the first conductance peak (n = 1), we obtain 0.185 meV we observe five MARs of n = 1,2,3,7 and 12 (see dashed lines in figure 4c). Interestingly, we observe a conductance peak at 0.44 mV that would indicate a slightly larger superconducting gap. However, as all other peaks do not match to this gap, we believe that this feature could be associated with the slightly higher $T_C$ of the polycrystalline Al contact pads. The first three Andreev peaks of the $V_G$ = -29.5 V slice align perfectly with the observed peaks at $V_G$ = 28 V. The stability of the MARs through 60 V of gate tuning (see supporting figure S4) further endorses the exceptional interface quality that we have achieved.

By fitting I/V characteristic curves above the Al superconducting gap ($V_D > 2\Delta/e$) (see supporting figure S5) we calculated the normal resistance ($R_n$) and the excess current ($I_{exc}$) for $V_G < 0$ V. $R_n$ converges at $V_G = -25$ V to 2.98 kΩ. Using the BTK[36] model we retrieve a barrier strength of $Z = 0.1$ resulting in a transparency of 99%, which is consistent with the transparency calculated from the quantized conductance plateaus.

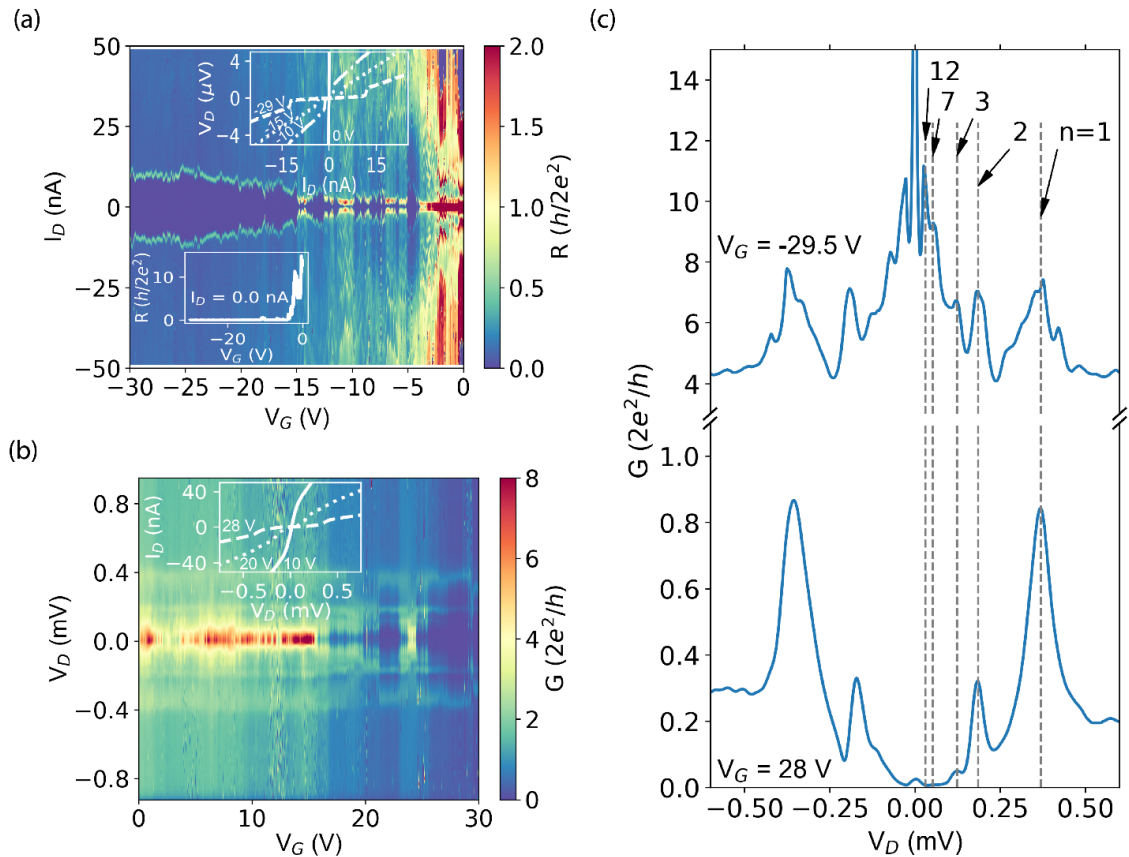

*Figure 4.* (a) Differential resistance $dV_D / dI_D$, with wiring resistance of 370 Ω subtracted, plotted in units of the quantum resistance versus $I_D$ and $V_G$. $I_D$ was swept from negative to positive and $V_G$ from 0 V to -30 V measured at 420 mK. The dark blue regions correspond to zero resistance and *indicate superconductivity induced into the Ge/Si core/shell channel. The upper inset shows $V_D$ versus $I_D$ for four different $V_G$ (0 V, -10 V, -15 V, -29 V). The lower inset shows a slice of differential resistance $dV_D / dI_D$ at $I_D$ = 0 nA with respect to $V_G$. (b) Differential conductance $dI_D / dV_D$, with wiring resistance of 370 $\Omega$ subtracted, plotted in units of quantum resistance versus $V_D$ and $V_G$. $V_D$ was swept from negative to positive. The inset shows $I_D$ versus $V_D$ for three different $V_G$ (10 V, 20 V, 28 V) (c) Differential conductance slice ($dI_D / dV_D$) with respect to $V_D$ for $V_G$ = -29.5 V and 28 V measured at 450 mK.*

In conclusion, we report the synthesis and electrical characterization of crystalline axial Al-Ge/Si-Al core/shell NW heterostructures. We have devised and implemented an annealing scheme to fabricate ultra-scaled Al-Ge/Si-Al heterostructures with highly transparent contacts. Through consecutive annealing cycles, the length of the Ge segment can be tuned beyond lithographic limitations. High-resolution HAADF STEM and EDX measurements show an abrupt atomically precise interface between the c-Al leads and Ge with an intact Si shell around the entire Al-Ge-Al heterostructure. The integration of these 1D hole gas NW heterostructures in back-gated FETs enabled the investigation of their transport properties. Observations of stable MARs and quantized conductance affirmed the great quality of the interface and a high contact transparency, greater than 98% of the metal-semiconductor contact. The observation of a supercurrent up to 15 nA at 450 mK and its ability to be tuned by a gate makes these devices suitable candidates for gatemon qubits. Further, the promising gap ratio of <$G_G$> / <$G_N$> = 0.03 in the low conductance regime is promising for the observation of Majorana zero modes. The opportunity to tune the electrical properties of these single crystalline metal-semiconductor heterostructures provides a promising platform towards practical applications of future ultra-scaled axial and radial nanoelectronic devices.

METHODS:

**Device fabrication:**

The starting materials are vapor-liquid-solid grown core/shell NWs with a Ge NW core diameter of 30 nm and a Si shell thickness of about 3 nm covered by a thin layer of native oxide.[8,4] The NWs were drop casted onto an oxidized highly p-doped Si substrate and the Ge core NW was contacted by Al pads fabricated by electron beam lithography, 100 nm Al sputter deposition and lift-off techniques. To gain access to the Ge core nanowire, the Si shell was selectively removed by wet chemical etching for 10 s in buffered HF (7:1) to remove the native oxide layer followed by a 15 s treatment in KOH (30%). The Al-Ge exchange reaction is induced by rapid thermal annealing at a temperature of T = 674 K in forming gas atmosphere and results in the substitution of the Ge core by c-Al.[12] The diameter of the Al core and the thickness of the Si-shell are thereby inherited from the initial Ge-Si core-shell NW geometry. Facilitating the proposed heterostructure formation scheme[13], allows the integration of the core-shell heterostructures with tunable channel lengths in a FET architecture, with the highly p-doped substrate acting as back-gate.

**Electrical characterization:**

The electrical measurements at room-temperature and ambient conditions were performed using a combination of a semiconductor analyzer (HP 4156B) and a probe station. To minimize the influence of ambient light as well as electromagnetic fields, the probe station is placed in a dark box. Low-temperature measurements (5 - 298 K) were performed in vacuum at a background pressure of approximately $2.5 \times 10^{-5}$ mbar using a $^4$He cryostat (Cryo Industries CRC-102) and a semiconductor analyzer (Keysight B1500A). Electrical measurements below T = 5 K were carried out using a self-built pumped $^3$He cryostat with a base temperature of T = 400 mK. Noise filtering was achieved using a room temperature Pi-filter and at low temperature thermal coax of approximately 1 m in length. The resistance of the fridge wiring was independently measured to be 370 Ω at 420 mK. The device was probed at low temperature using both voltage and current biasing techniques with a National Instruments PCI DAC/ADC high frequency card. In the voltage biasing scheme a voltage divider consisting of a ratio of 50kΩ/50Ω was used to reduce the amplitude of the voltage source. A Femto variable gain transimpedance amplifier (DCPCA-200) was used to convert and amplify the induced current to a voltage signal measured by the NI card. In the current biasing scheme a 10 MΩ resistor was used to convert the voltage signal to a current signal with a maximum amplitude of 1 μA. The current was applied to the sample which was grounded at one end. The potential difference across the sample was amplified by two NF Electronic Instruments low noise preamplifiers (LI-75A), each of a gain of 100, in series. The back-gate was biased using a Yokogawa programmable voltage source.

AUTHOR INFORMATION

Corresponding Authors

#Email: olivier.buisson@neel.cnrs.fr

Author Contributions

M.S. performed the device fabrication, experimental design. M.S., J.D., C.N., and R.K. conducted the electrical characterisation. A.L., O.B., C.N. and J.S. conceived the project, contributed essentially to the experimental design and provided expertise on theoretical interpretations. M.I.H., L.M.A. and E.R. carried out the TEM and EDX measurements and analysis. J.Y. and C.M.L provided the Ge/Si core/shell NWs, helpful feedback and commented on the manuscript. All authors analysed the results and helped shape the research and manuscript.

Notes

The authors declare no competing financial interest.

ACKNOWLEDGMENT

The authors gratefully acknowledge financial support by the Austrian Science Fund (FWF): project No. P29729-N27. The authors further thank the Center for Micro- and Nanostructures for providing the cleanroom facilities. We acknowledge support from the Laboratoire d'excellence LANEF in Grenoble (ANR-10-LABX-51-01). Financial support from the ANR-COSMOS (ANR-12-JS10-0002) project is acknowledged. We acknowledge support from Campus France in the framework of PHC AMADEUS 2016 for PROJET N° 35592PB. J. Delaforce acknowledges the European Union's Horizon 2020 research and innovation programme under the Marie Skłodowska-Curie grant agreement No 754303. We benefitted from the access to the Nano characterization platform (PFNC) in CEA Minatec Grenoble. The authors would like to acknowledge Silvano De Franceschi, François Lefloch, Kazi Rafsanjani and Tom Vethaak for beneficial discussions.